\documentclass[fleqn,10pt]{wlscirep}
\usepackage[utf8]{inputenc}
\usepackage[T1]{fontenc}
\usepackage{graphicx}
\usepackage{hyperref}
\usepackage[skip=0.333\baselineskip]{caption}
\usepackage{subcaption}
\title{Non-Markovian Quantum State Diffusion for the tunnelling in SARS-COVID-19 Virus }

\author[1]{Muhammad Waqas  Haseeb}
\author[2,*]{Mohamad Toutounji}

\affil[1]{Department of Physics, United Arab Emirates University, Al-Ain, UAE}
\affil[2]{Department of Chemistry, United Arab Emirates University, Al-Ain, UAE}

\affil[*]{mtoutounji@uaeu.ac.ae}


\begin{abstract}
In the context of biology, unlike the well-established Standard Model in physics, many biological processes lack a complete theoretical framework and are often described phenomenologically. A pertinent example is olfaction, the process through which humans and animals distinguish various odors. The conventional biological explanation for olfaction relies on the lock and key model, which, while useful, does not fully account for all observed phenomena. As an alternative or complement to this model, vibration-assisted electron tunnelling has been proposed.
Drawing inspiration from the vibration-assisted electron tunnelling model for olfaction, we have developed a theoretical model for electron tunnelling in SARS-CoV-2 virus infection within a non-Markovian framework. We approach this by solving the non-Markovian quantum stochastic Schrödinger equation. In our model, the spike protein and the GPCR receptor are conceptualized as a dimer, utilizing the spin-Boson model to facilitate the description of electron tunnelling. Our analysis shows that electron tunnelling persists even at intermediate and strong coupling limits between the dimer components, presenting a stark contrast to the predictions from Markovian regime models. Notably, Markovian models often yield unphysical negative probabilities, particularly in the strong coupling limit, underscoring significant discrepancies and highlighting the importance of considering non-Markovian dynamics in accurately modeling such quantum processes. \textcolor{black}{This approach enhances our understanding of viral infection mechanisms while also offering deeper insights into the quantum biological process of olfaction.}

\end{abstract}
\begin{document}

\flushbottom
\maketitle
\thispagestyle{empty}

%\noindent Please note: Abbreviations should be introduced at the first mention in the main text – no abbreviations lists. Suggested structure of main text (not enforced) is provided below.

\section*{Introduction}
\label{sec:introduction}
Quantum mechanics has traditionally been the domain of physicists, focusing on the fundamental particles and forces that govern the universe \cite{sakurai2020modern}. However, the field has expanded significantly. Its principles have found applications far beyond the microscopic interactions of inanimate matter, penetrating the realms of chemistry and biology. One such frontier is the exploration of quantum phenomena in biological systems, an area that promises to redefine our understanding of life's fundamental processes\cite{mohseni2014quantum,mcfadden2018origins,engel2007evidence,van2011quantum,alkobaisi2025artificial}.\\ The fundamental senses of hearing, sight, touch, taste, and smell are essential for both animals and humans to interact with their environment. Among these, the sense of smell, or olfaction, is particularly vital as it enables the detection and differentiation of various scents through volatile odorant molecules 
 \cite{goldstein2009encyclopedia}. This ability is crucial for identifying food sources and avoiding predators. While the basic mechanisms of most senses are well understood, the underlying processes of olfaction remain a subject of ongoing scientific debate \cite{turin1996spectroscopic,brookes2007could}. The prevalent lock-and-key model suggests that the interaction of odorants with olfactory receptors, which are a type of G-protein coupled receptor (GPCR), triggers olfaction \cite{scott2001chemosensory}. According to this model, the physical and chemical properties of the odorant molecules—such as size, shape, and functional groups—determine the activation of these receptors, leading to a neural response. Recent studies challenge the traditional lock-and-key model of olfaction, which suggests that molecular shape determines receptor activation. Research involving benzaldehyde isotopes indicated that humans and Drosophila can distinguish isotopes based on vibrational mode  differences, not just molecular shape \cite{haffenden2001investigation,bittner2012quantum}. This supports the hypothesis that olfaction may involve sensing vibrational spectra through electron transfer in olfactory receptors, suggesting a more complex mechanism of smell perception than previously thought \cite{turin1996spectroscopic,franco2011molecular,hoehn2015neuroreceptor,yu2007tornado}.\\ 
The emergence of SARS-CoV-2 has not only reshaped our global landscape but also hinted at the onset of an era characterized by frequent pandemics \cite{peters2022coming}. This situation underlines the critical need to fast-track research into the mechanisms of viruses to better understand and combat their effects. Exploring interdisciplinary approaches could be particularly beneficial, offering novel insights into the complex physiological interactions of viruses with their hosts, potentially leading to breakthroughs in prevention and treatment strategies  \cite{mohseni2014quantum}. Traditional virology has provided substantial insights into the mechanisms of viral infection and replication 
 \cite{belouzard2012mechanisms,lan2020structure,alkobaisi2012interactive}. Yet, the complete dynamical landscape of how viruses interact with host cells at the quantum level remains  unexplored \cite{gray2003electron,ahmed2017open}. The spike protein of SARS-CoV-2 and its interaction with the human ACE2 receptor is a critical step in the virus's mechanism of cellular entry \cite{hippisley2020risk,abd2020human,rosenbaum2009structure}. Understanding this interaction down to the quantum level could open up new pathways for therapeutic interventions and preventive measures.
 \textcolor{black}{In the vibration-assisted electron tunnelling model, odorant recognition is proposed to occur via inelastic electron tunnelling processes across a binding site in olfactory receptor proteins\cite{turin1996spectroscopic,brookes2007could,solov2012vibrationally}. When an odorant molecule binds, its vibrational modes interact with electrons tunnelling between donor and acceptor sites within the receptor. If the vibrational energy matches the energy difference between these sites, tunnelling is facilitated, initiating downstream signal transduction. This mechanism allows receptors to function as molecular spectrometers sensitive to specific vibrational energies.
Compared to the classical "lock-and-key" model, which relies purely on molecular shape, the vibrational model offers a crucial advantage: it can explain isotope effects. Specifically, molecules such as acetophenone and its fully deuterated analogue (acetophenone-d$_8$), which are structurally identical are perceived differently in odor, a phenomenon shape-based theories cannot account for. In contrast, the vibrational model predicts that isotope substitution shifts vibrational frequencies (e.g., C--H stretch to C--D stretch), thereby altering tunnelling probabilities and perceived odor\cite{turin1996spectroscopic}. Furthermore, recent studies on freshwater planarians have demonstrated the isotope effect in a biological context. Notably, these studies have observed a distinct difference in the locomotion patterns of freshwater planarians when exposed to cotinine and nicotine\cite{kakuturu2024schild,ruble2024cotinine}. The difference in behavioral response is attributed to the isotopic variation between these molecules, suggesting that vibrational differences, rather than mere structural similarity, play a significant role in modulating biological activity. These findings support the hypothesis that vibrationally assisted mechanisms, including electron tunnelling, may be relevant in complex biological processes, further aligning with the theoretical framework proposed in our study.
Regarding the specific coupling mechanism, Turin proposed that olfactory receptors (GPCRs) contain redox active sites involving NADPH electron donors and zinc dependent acceptor motifs. Binding of the odorant enables electron tunnelling by providing vibrational energy matching, ultimately leading to reduction of disulfide bonds and activation of G-protein signaling pathways.
Although direct experimental proof of vibration-assisted tunnelling in olfaction remains under investigation, this includes indirect evidence such as differences. These include differences in odor perception between isotopologues, neurophysiological studies showing odorant response patterns correlating with vibrational features rather than structural features \cite{rosenbaum2009structure,lechien2020olfactory,bae2007interactive}, and biochemical identification of zinc-binding and NADPH binding motifs in olfactory receptors. Therefore, we believe this quantum biological model provides a compelling and testable explanation for molecular recognition in olfaction.}
 Drawing on the concept of vibration-assisted electron tunnelling, which plays a pivotal role in the olfaction process, we propose a similar mechanism of electron tunnelling that may occur during the COVID-19 infection process. This process could potentially act as a molecular switch, augmenting the traditional lock-and-key model of molecular interaction. By incorporating this quantum mechanical perspective, we aim to enrich the existing understanding of how the SARS-CoV-2 virus triggers cellular mechanisms, potentially offering new avenues for therapeutic intervention.\\
 This paper is structured into five main sections, each addressing a critical aspect of the research. Section 1, titled "Introduction," provides an overview of the research context, highlights its significance, and briefly reviews relevant prior work. Section 2, titled "Non-Markovian Quantum State Diffusion," elaborates on the research methodology, including data collection techniques, analytical methods employed, and the limitations associated with the chosen approach.
Section 3 focuses on the modeling of virus infection and the derivation of the Stochastic Schrödinger Equation (SSE), providing the theoretical foundation of our study. Section 4, titled "Results and Discussion," presents the primary findings, accompanied by a thorough analysis and interpretation. The results are compared with previous studies to provide a comprehensive perspective.
Finally, Section 5, "Conclusion," encapsulates the key outcomes, discusses their broader implications, and suggests potential directions for future research.
  \begin{figure}[ht!]
\centering
\includegraphics[width=14cm]{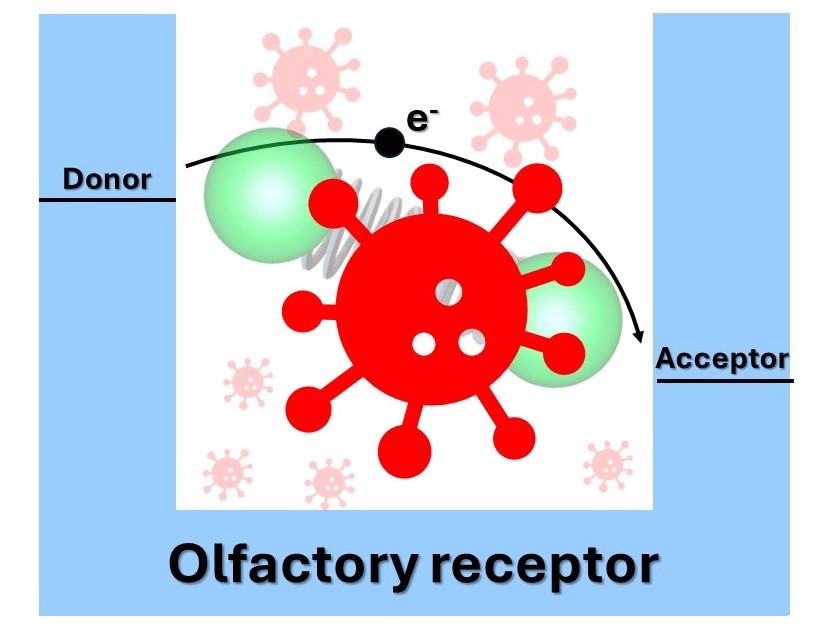}% Here is how to import EPS art
\caption{Provides a simplified graphical representation illustrating the process of intra-protein electron transfer that occurs during the interaction between the SARS-CoV-2 virus and a host cell receptor. It visually captures the critical molecular events that facilitate the transfer of electrons within the proteins involved, crucial for the viral attachment and entry mechanism into the host cells.}
\label{fig-1}
\end{figure}
\section*{A quantum approach to biological processes}
Quantum tunnelling has been a key concept in quantum biology, offering insights into fundamental biological processes such as DNA mutations and enzyme catalysis.\cite{mcfadden2018origins} In the 1960s, Löwdin proposed that proton tunnelling could be the mechanism behind spontaneous mutations in DNA, an idea that continues to receive attention in the scientific community. \cite{outeiral2021prospects} Additionally, enzymes, traditionally understood through the lock-and-key model,\cite{fischer1894einfluss} were also suggested to exploit quantum tunnelling for both electron and proton transfer.\cite{de1966studies} This mechanism has since been substantiated through extensive research and is now a well-established field of study.\cite{bothma2010role,moser2010guidelines} Receptor-ligand binding, particularly involving G-protein coupled receptors (GPCRs), also plays a significant role in various physiological processes such as olfaction and neurotransmitter action.\cite{uings2000cell} GPCRs, including rhodopsins that bind chromophores involved in light detection, play a central role in photosynthesis, where quantum coherence facilitates both energy and charge transfer. \cite{engel2007evidence,brixner2005two} These biological phenomena reveal a complex interplay between vibrational and electronic states, often facilitated by proteins, which serve to enhance the efficiency of charge transfer processes,\cite{solov2012vibrationally}

In this context, the vibrational theory of olfaction proposes that receptor activation is driven not by molecular shape, but by a molecule's vibrational spectrum, which facilitates electron tunnelling between olfactory receptors and odorant molecules\cite{turin1996spectroscopic}. Although some experimental evidence, such as the ability to differentiate deuterated odorants,\cite{franco2011molecular,hoehn2015neuroreceptor} supports this theory, there remains some scepticism with regards to theory. \cite{block2015implausibility} Recent  research has extended this quantum tunnelling framework to neurotransmitter binding, a process crucial for neural signal transmission. The vibrational spectra associated with specific neurotransmitters like serotonin, histamine, and adenosine could potentially facilitate charge transfer during neurotransmission, although direct experimental confirmation is still needed.\cite{hoehn2017experimental,chee2015characteristic} The broader field of quantum biology has long recognized charge transfer within biological systems, particularly in membrane-embedded proteins, where environmental factors such as protein vibrations or ligand binding play a critical role.

\textcolor{black}{In the case of SARS-CoV-2, quantum tunnelling has also been suggested as a potential mechanism in the virus's interaction with host cells. The virus invades cells by binding its spike protein to the angiotensin-converting enzyme 2 (ACE2) receptor, which is embedded in the host cell membrane.\cite{ni2020role,bhalla2021historical}ACE2 plays an essential role in cardiovascular function, as it modulates the hormone angiotensin. Mutations in the spike protein that alter its interaction with ACE2 have raised concerns about increased viral transmissibility. Beyond ACE2, other enzymes, such as serine protease TMPRSS2 and cathepsin L, are involved in preparing the spike protein for cell entry, further emphasizing the role of enzymes in the viral infection process.\cite{seth2020covid,hoffmann2020sars} Given the established role of quantum tunnelling in enzyme function, it seems plausible that quantum effects may also influence these interactions.
The vibration-assisted tunnelling model could plausibly be extended to other sensory modalities involving GPCR activation, including taste and neurotransmission. TAS2Rs (bitter taste receptors) are GPCRs structurally similar to olfactory receptors.\cite{hoehn2015neuroreceptor} Like olfactory receptors, they could support electron tunnelling facilitated by vibrational coupling of bound ligands. Bitter molecules often contain functional groups (e.g., hydroxyl, amine, or aromatic rings) that provide distinct vibrational fingerprints, making them strong candidates for vibration-assisted detection.
Moreover, vibrational mechanisms have been discussed in neurotransmitter receptor interactions. For example, molecules like serotonin and dopamine involve aromatic residues and hydrogen bonding, suggesting that vibrationally modulated electron tunnelling could contribute to GPCR activation in synaptic signaling.\cite{adams2020quantum} Although experimental validation in taste or neurotransmission is still in early stages, the theoretical framework supports the generalization of vibration-assisted tunnelling beyond olfaction.}
There is growing interest in how electron transfer, facilitated by protein vibrations, might play a role in the binding mechanisms of SARS-CoV-2, particularly through disulfide bridges within ACE2 and the spike protein.\cite{hati2020impact} Both proteins are rich in cysteine residues, and evidence suggests that their interaction relies on the redox potential of these disulfide bridges, which influences viral infectivity.\cite{singh2020sars} This redox potential has been shown to affect binding affinity, with disruption of the disulfide bonds significantly impairing the interaction between the spike protein and ACE2. Although the ACE2 receptor is not a GPCR, its structural similarities with certain GPCRs—particularly the presence of disulfide bridges—warrant further investigation into the possibility of electron transfer in these receptors.\cite{lechien2020olfactory} Moreover, the role of GPCRs in SARS-CoV-2 infection, particularly in olfaction and the cytokine storm response, points to the broader relevance of quantum biological processes in understanding viral infections.\cite{hojyo2020covid}
\section*{Non-Markovian Quantum State Diffusion}
Quantum systems in biological environments do not operate in isolation; they interact continually with a myriad of other biological components that constitute a complex, noisy background. It's vital to account for its interaction with the surrounding environment or reservoir. Such incorporation becomes feasible through the open quantum system (OQS) framework. 
 \cite{breuer2002theory} This framework encapsulates the dynamics of quantum systems by integrating the Hamiltonian of the primary system with that of the external environment and their mutual interactions. A prevalent technique to explore these dynamics is through the use of master equations, such as the Nakajima-Zwanzig equation. \cite{nakajima1958quantum} Redfield's master equation, \cite{redfield1965theory} the Lindblad formula, \cite{manzano2020short} Hu-Paz-Zhang equation, \cite{ford2001exact} and the broad non-Markovian time-local master equation. \cite{zhang2019exact} These equations focus on the central system’s reduced density matrix, obtained by tracing over the reservoir’s degrees of freedom, and allow for the computation of time-evolving expectation values for any operator within the system’s Hilbert space. While traditional methods like Fermi’s Golden Rule provide a basis for calculating transition rates in systems with weak electronic coupling, recent studies  suggest these semi-classical theories fall short, especially in low-frequency environments. \cite{chkecinska2015dissipation} For instance, the use of a polaron master-equation approach has demonstrated that incorporating strong dissipation not only addresses discrepancies in classical rate predictions but also significantly enhances the system’s switching capabilities, offering greater selectivity for specific odorant modes despite an increase in background noise. This revelation underscores the limitations of semi-classical theories and highlights the potential benefits of advancing beyond classical constraints to improve sensory precision in molecular systems.

The Quantum State Diffusion (QSD) presents a robust alternative for analyzing the dynamics of open quantum systems (OQS), highlighting its advantages over traditional Master equation approaches, particularly in handling a bosonic environments. As detailed by Bouten et al. (2004), the QSD offers a comprehensive depiction of the system’s reduced density matrix without approximations, by tracing out environmental degrees of freedom and representing the dynamics through an ensemble of continuous trajectories driven by a complex Gaussian process. \cite{bouten2004stochastic} This methodology excels by allowing expansions based on the environmental correlation time relative to the system's typical timescale. Initially, this leads to the Markovian approximation at the expansion's leading order, but it is the subsequent orders that furnish the non-linear non-Markovian  Stochastic Schrödinger Equation (SSE), capturing more complex dynamics that the Markovian model cannot. This approach is particularly effective in strong coupling scenarios and transitions seamlessly from Markovian to non-Markovian regimes, enhancing its utility in practical applications. Furthermore, SSE has proven its efficacy and efficiency in numerical simulations, offering results that align well with those derived from the Lindblad-form Master equations for Markovian systems, thereby affirming its accuracy and reliability in quantum dynamics studies. \cite{yu1999non}\\
Our study extends the non-Markovian Quantum State Diffusion (NMQSD) framework to investigate the role of electron tunnelling  between the SARS-CoV-2 spike protein and the ACE2 receptor.  We model the spike protein and the ACE2 receptor as a quantum system interacting with a biological environment (membrane), using a spin-Boson model to represent the two-level systems and their interactions with an external bath. This model allows exploration of how spike protein vibrations may facilitate electron tunnelling at the receptor interface, potentially enhancing viral entry efficiency. This research not only deepens our understanding of viral mechanisms but also bridges quantum physics with biological sciences.

A quantum system coupled to a bosonic bath is studied, described by the Hamiltonian \cite{yu1999non} (with \(\hbar = 1\)):
\begin{equation}
\label{eq-1}
H = H_{\text{sys}} + \sum_k (g_k L b_k^\dagger + g_k^* L^\dagger b_k) + \sum_k \omega_k b_k^\dagger b_k,
\end{equation}
where \(H_{\text{sys}}\) denotes the system Hamiltonian, \(L\) the Lindblad operator, \(b_k\) the annihilation operator of the k-th mode of the bath with frequency \(\omega_k\), and \(g_k\) the coupling strength. The bath correlation function, central to the system's dynamics, is expressed as \cite{strunz1999open}:
\begin{equation}
\label{eq-2}
\alpha(t,s) = \sum_k |g_k|^2 e^{-i\omega_k (t-s)}.
\end{equation}

By defining $z_t^* \equiv -i \sum_k g_k^* z_k^* e^{i \omega_k t}$ as a function characterizing the time-dependent states of the bath, and considering \( z_k \) as Gaussian random variables, \( z_t^* \) becomes a Gaussian random process with zero mean \( M[z_t^*] = 0 \) and the correlation function \( M[z_t z_s^*] = \alpha(t,s) \), where \( M[\cdot] \) denotes the ensemble average.

At zero temperature, the system's quantum trajectory obeys a linear, time-local Quantum State Diffusion (QSD) equation:
\begin{equation}
\label{eq-3}
\frac{\partial}{\partial t} |\psi_{z^*}(t)\rangle = [-iH_{\text{sys}} + Lz_t^* - L^\dagger \bar{O}(t,z^*)] |\psi_{z^*}(t)\rangle
\end{equation}
where \( O \) is an operator ansatz defined by the functional derivative
\begin{equation}
\label{eq-4}
\frac{\delta}{\delta z_s^*} |\psi_{z^*}(t)\rangle = O(t,s,z^*) |\psi_{z^*}(t)\rangle,
\end{equation}
and $\overline{O}(t,z^*) = \int_0^t \alpha(t,s)O(t,s,z^*)\,ds$.\\
The system's reduced density operator \( \rho_s(t) \) can be calculated through ensemble averages of quantum trajectories.\\
\begin{equation}
\label{eq-5}
\rho_s \equiv \mathrm{Tr}_ {\mathrm{env}} \left[ e^{-iHt} |\psi_0\rangle \langle \psi_0| \otimes \rho_{\mathrm{env},0} e^{iHt} \right] = \mathcal{M} \left[|\psi_t(z)\rangle \langle \psi_t(z)|\right] .
\end{equation}
 The QSD approach involves deriving the functional derivative \( O \) operator, which can be exact for simple models or perturbatively derived for more general systems.

The non-Markovian unraveling of Quantum State Diffusion (QSD) predicated on states with normalization
\begin{equation}
\label{eq-6}
\tilde{\psi}_t(z) = \frac{\psi_t(z)}{\|\psi_t(z)\|}
\end{equation}
could be achieved using the Girsanov transformation\cite{diosi1997non}.
\begin{multline}
\label{eq-7}
\frac{d}{dt}\tilde{\psi}_t = -iH_{sys}\tilde{\psi}_t + (L - \langle L\rangle_t)\tilde{\psi}_t \tilde{z}_{t} -
\int_0^t ds \, \alpha(t,s) \left\langle (L^\dagger - \langle L^\dagger\rangle_{s}) \hat{O}(t,s,{\tilde{z}_t}) - (L^\dagger - \langle L^\dagger\rangle_{s})\hat{O}(t,s,{\tilde{z}_t}) \right\rangle \tilde{\psi}_t
\end{multline}
Where ${\tilde{z}_t}$ is the shifted noise, 
\begin{equation}
\label{eq-8}
{\tilde{z}_t}=z_t+\int_0^t ds \, \alpha(t,s) \langle L^\dagger\rangle_{s}
\end{equation}
and \( \langle L\rangle_{s}=\langle \tilde{\psi}_t| L |\tilde{\psi}_t \rangle \) is the quantum average. 

The non-linear non-Markovian QSD equation can be written in a compact form by introducing \( \Delta_t(A)=A-\langle A\rangle_t \)
\begin{equation}
  \label{eq-9}
\frac{d}{dt}\tilde{\psi}_t = -iH_S\tilde{\psi}_t + \Delta_t(L)\tilde{\psi}_{t}z^{\ast}_t - \Delta_t(L^{\dagger})\bar{O}(t, \tilde{z})\tilde{\psi}_t + \Delta_t(L^{\dagger})\bar{O}(t, \tilde{z})_t\tilde{\psi}_t  
\end{equation}

Where 
\begin{equation}
\label{eq-10}
\bar{O}(t, z) = \int_{0}^{t} ds \, \alpha(t,s)\hat{O}(t,s,z)
\end{equation}
The (Eq.~\ref{eq-7} or \ref{eq-9}) is fundamental for non-Markovian Quantum state diffusion and the perturbative treatment starts with this equation. Applying the formal Perturbation theory on operator $\hat{O}(t, s, z)$ using a series expansion in powers of $(t-s)$. \cite{strunz1999open}.
\begin{multline}
\label{eq-11}
\frac{d}{dt}\tilde{\psi}_t = -iH_{Sys}\tilde{\psi}_t + \Delta_t(L)\tilde{z}_t - g_0(t)((\Delta_t(L^{\dagger})L - \langle\Delta_t(L^{\dagger})L \rangle_t))\tilde{\psi}_t + ig_1(t)(\Delta_t(L^{\dagger})[H,L] - \langle\Delta_t(L^{\dagger})[H,L]\rangle_t))\tilde{\psi}_t\\
+ g_2(t)((\Delta_t(L^{\dagger})[L^{\dagger},L]L - \langle\Delta_t(L^{\dagger})[L^{\dagger},L]\rangle_t))\tilde{\psi}_t
\end{multline}
In the study of non-Markovian Quantum State Diffusion, the first-order corrections to the zeroth-order term depend on the system's typical frequencies and relaxation rates, symbolized by \(\omega\) and \(\Gamma\) respectively. These corrections become relevant when environmental correlation times are finite but short compared to the system's timescales, ensuring the expanded QSD equation's applicability. As the correlation time  approaches zero, the dynamics transition into the Markovian regime, where only the zeroth-order term remains, thus reverting to the conventional Markov QSD equation. This highlights the conditions determining whether a quantum system will exhibit non-Markovian or Markovian dynamics.\\

We choose the spectrum of the environment to be of Lorentz-type, described by the spectral density function
\begin{equation}
\label{eq-12}
J(\omega) = \frac{\Gamma}{\pi} \frac{\gamma^2}{\omega ^2 + \gamma^2}, 
\end{equation}
where $\gamma^{-1}$ is the environmental memory time which corresponds to the cutoff frequency $\omega_c$.
In the Markovian regime, the system's coupling strength \(\Delta\) is much smaller than the bath's cutoff frequency \(\omega_c\), i.e., \(\Delta \ll \omega_c\). This results in a bath that evolves much faster than the system, losing its memory quickly. Consequently, the system behaves as if each interaction with the bath is independent of the previous ones, making the Markovian approximation appropriate. This situation is typical in many physical systems, such as ion traps, quantum dots, and superconducting devices, where the fast bath dynamics dominate.
It is qualitatively different in the bio-molecular environments where,    the system's coupling strength \(\Delta\) becomes comparable to or greater than the bath's cutoff frequency, i.e., \(\Delta \geq  \omega_c\), while the bath responds more slowly compared to the system’s evolution. This results in the bath retaining memory of its interactions with the system, which leads to more complex dynamics that cannot be described by Markovian approximations. Such non-Markovian behavior is common in biological systems  dynamics in biomolecular environments, where \(\hbar \omega_c\) is typically on the order of 2-8 meV, and \(\hbar \Delta\) ranges from 0.2 meV to 100 meV. In this regime, coherent oscillations and memory effects dominate the system's evolution, requiring more sophisticated approaches to capture the dynamics accurately \cite{thorwart2009enhanced}.\\
In the context viral of the infection mechanism at the molecular level, using the Markovian approximation can result in the prediction of unphysical negative probabilities of electron tunnelling , especially at the highest coupling strength\cite{adams2022quantum,haseeb2024vibration}. These results are not physically viable, revealing the inadequacies of the Markovian approach in accurately modeling such complex interactions. To resolve this, our model considers the specific parameters of SARS-CoV-2 infection and its bio-molecular environment, allowing for a relaxation of the Markovian approximation. By adopting a non-Markovian framework, we incorporate memory effects into the system-environment interaction, which leads to more accurate and physically consistent predictions. This approach not only eliminates the unphysical negative probabilities but also provides a more refined understanding of how electron tunnelling contributes to the interaction between the SARS-CoV-2 spike protein and the ACE2 receptor.  Although Eq. \ref{eq-11} is formally derived under the zero-temperature approximation, the influence of finite temperature can be systematically incorporated into the model through the time-dependent coupling constants $g_0(t)$, $g_1(t)$, $g_2(t)$, which encode the temporal structure of the bath’s thermal fluctuations via their relationship to the bath correlation functions. At finite temperatures, these coupling constants acquire temperature dependence through the occupation factors of the bath modes, governed by the Bose-Einstein distribution, and the spectral density of the bath. Specifically, the time-correlation functions of the bath, which describe the memory effects of the environment, become thermally broadened, reflecting the interplay between quantum coherence and thermal noise.
Assuming $\Gamma = 1 $,  and considering that the noise is of a complex Ornstein–Uhlenbeck type, the correlation function can be expressed as \cite{gribben2020exact}
\begin{equation}
\label{eq-13}
\alpha(t,s) = \int_0^\infty d\omega J(\omega) [\coth\left(\frac{\omega}{2 k_B T}\right) (\cos(\omega t) - i\sin(\omega t))]
\end{equation}

The Quantum State Diffusion (QSD) equation is central to understanding quantum dynamics, particularly how a system interacts with its environment. This equation incorporates the system's Hamiltonian, the Lindblad operator, and their commutators. The key to applying the QSD equation lies in determining the coefficients \( g_i(t) \), which are directly influenced by the environmental correlation function \( \alpha(t,s) \). These coefficients capture the essence of the quantum noise and are particularly distinct under Ornstein-Uhlenbeck noise with its exponential correlation. Understanding \( g_i(t) \) is crucial for leveraging the QSD equation in practical scenarios involving quantum systems affected by external noise.These coefficients can be computed as. 
\begin{equation}
\label{eq-14}
g_0(t) = \int_0^t \alpha(t,s) ds
\end{equation}

\begin{equation}
\label{eq-15}
g_1(t) = \int_{0}^{t} \alpha(t,s)(t-s)ds
\end{equation}
\begin{equation}
\label{eq-16}
g_2(t) = \int_0^t du\int_0^s \alpha(t,s) \alpha(s,u) (t-s)  ds
\end{equation}
The electron tunnelling regime in this system is governed by the interplay between thermal energy ($k_B T = 0.025$ eV), vibrational frequencies of virus ($ \omega_i = 1040,0.1240, 0.1570$ eV), and system-bath coupling ($ \gamma_{iE} = 0.005$ eV).These parameters are detailed in Table 1. At room temperature ($T \approx 290$ K), the thermal energy $k_B T$ is significantly smaller than the vibrational mode of the virus $ \omega$, placing the system in a quantum regime where thermal fluctuations do not significantly populate higher vibrational states.
For the given parameters ($k_B T = 0.025\,\text{eV}$, $ \omega = 0.1240\,\text{eV}$, $ \gamma_{iE} = 0.005\,\text{eV}$), the hierarchy $k_B T \ll \omega$ ensures that thermal energy remains insufficient to populate higher vibrational states, preserving the quantum mechanical nature of the vibrational mode. However, the weak system-bath coupling ($ \gamma_{iE} \ll k_B T$) implies that thermal fluctuations only perturb the system’s dynamics, allowing the bath correlation functions to retain non-Markovian features over timescales relevant to electron tunnelling. In this highly quantum regime, the electron transfer is dominated by the discrete structure of the vibrational mode, and the temperature has limited effect on the transfer rate.\cite{so2024trapped} \textcolor{black}{ Experimental signatures of non-Markovian dynamics involve observable memory effects, wherein the system's evolution depends on its past history. In practice, single-molecule fluorescence experiments can detect non-Markovianity by analyzing blinking statistics or fluorescence intermittency patterns that deviate from simple exponential decay. \cite{kondo2017single} Additionally, Raman spectroscopic techniques, particularly 2D-Raman and time-resolved Raman spectroscopy, are capable of revealing coherence lifetimes and vibrational memory signatures. The observation of extended coherence times and revival phenomena in such measurements are consistent with and indicative of non-Markovian behavior \cite{lukose2023raman}.}

The temperature dependence of $g_0(t)$ $g_1(t)$,and $g_2(t)$ thus manifests as a modulation of the tunnelling rate, where thermal noise introduces stochasticity in the tunnelling pathway while quantum coherence mediated by vibrational modes persists. This dual behavior arises because the bath correlation time, dictated by the inverse of the bath spectral width and temperature, remains comparable to the timescale of coherent tunnelling processes. Consequently, even at finite temperatures, the system retains memory of its quantum state, enabling vibration-assisted tunnelling to proceed efficiently despite thermal perturbations. 

This framework reconciles the zero-temperature derivation with experimental conditions (e.g. room temperature) by explicitly embedding thermal effects into the non-Markovian dynamics, avoiding the unphysical predictions of Markovian models (e.g., negative probabilities in strong coupling) that neglect memory effects. The result is a robust description of electron tunnelling in regimes where quantum coherence and thermal fluctuations coexist, as observed in biological systems such as viral spike protein-receptor interactions, where non-Markovian quantum effects are critical to mechanistic accuracy.\\

\section*{Model}
Biological phenomena are often studied through the lens of open quantum systems, taking into account interactions between the system and its surrounding environment \cite{mohseni2014quantum}. Electron transfer processes, particularly those assisted by vibrational mechanisms, are typically analyzed using the established spin boson model \cite{gilmore2005spin, chkecinska2015dissipation}. This model represents the system as a donor-acceptor duo linked to an environmental bath, which is effectively modeled as a collection of harmonic oscillators \cite{marais2013decoherence}. In our approach, the entire complex comprising the ligand protein and its receptor, and their immediate environment is considered as a unified system. Inspired by the mechanisms proposed for olfaction \cite{solov2012vibrationally}, we utilize the framework of open quantum systems to investigate the relationship between the peak transfer probability at the ACE2 receptor and its interaction with a vibrational mode of the SARS-CoV-2 spike protein. To facilitate analysis, the receptor system is modeled as a dimer, and the principal Hamiltonian delineates the dynamics and interactions within this system and its surrounding environment. A schematic depicting the electron transfer process within the protein is presented in Figure~\ref{fig-1}.
\begin{figure}[h!]
\centering
\includegraphics[width=18cm]{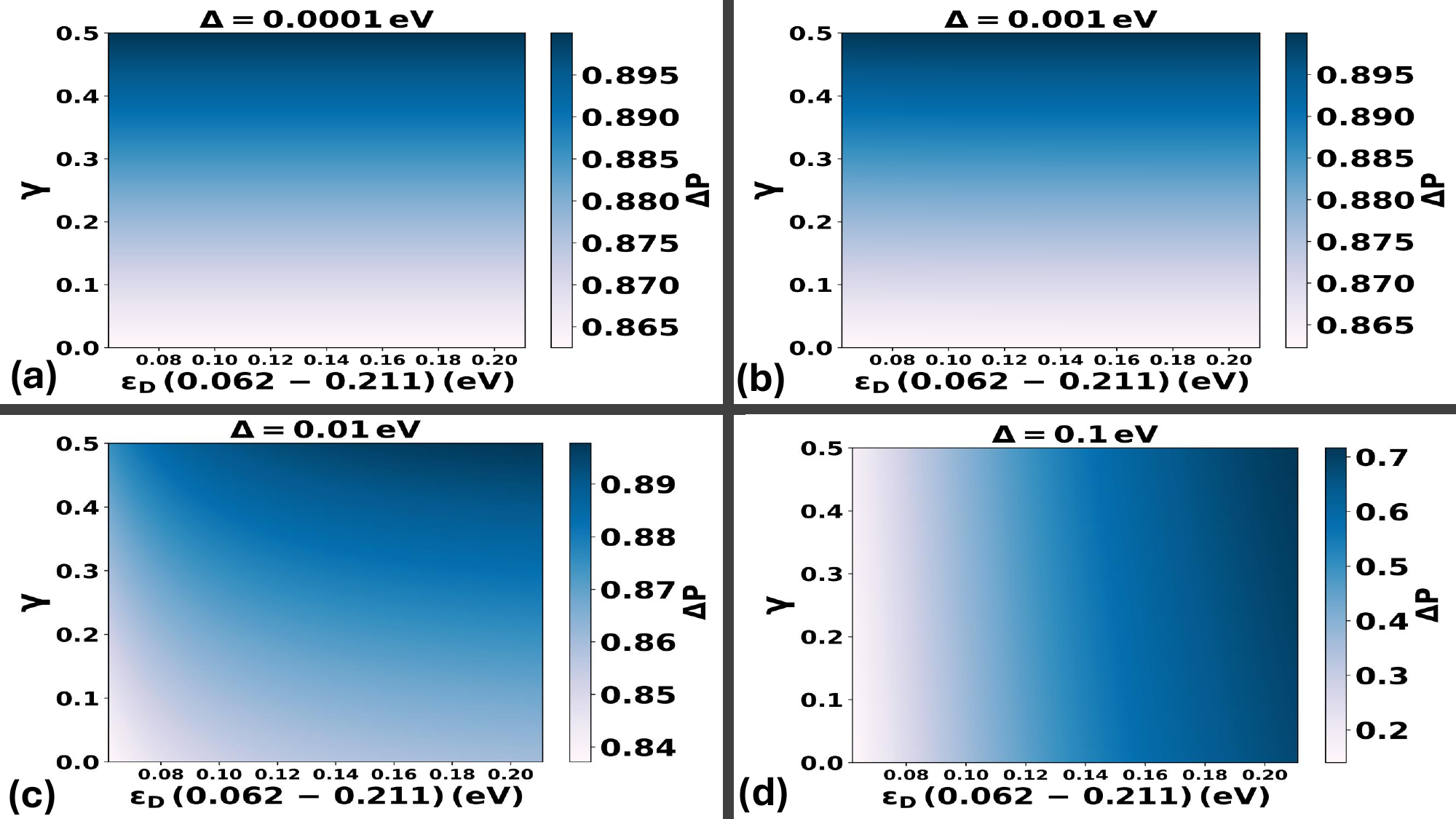}% Here is how to import EPS art
\caption{ (a-d) The influence of the vibrational mode  of the spike protein on electron transfer probabilities across different dimer coupling strengths is illustrated in (a)-(d). (a) shows the scenario at $\Delta = 0.0001 eV$, representing the weak coupling limit, where the impact of the vibrational mode   on transfer probability is minimal. In  (b) at $\Delta = 0.001 eV$, a slightly stronger coupling results in a small increase in transfer probability. (c)at $\Delta = 0.01 eV$ where the influence of the vibrational mode   on transfer probability becomes more pronounced. Finally, (d) at $\Delta = 0.1 eV$ in the strong coupling limit, shows a significant enhancement in transfer probability due to the vibrational mode  , highlighted by the white region. }
\label{fig-2}
\end{figure}
The Hamiltonian for the receptor is 
\begin{equation}
\label{eq-17}
H_{R} = \frac{1}{2} \epsilon \sigma_z + \frac{1}{2} \Delta \sigma_x
\end{equation}
For a dimer isolated from external interactions, the maximum probability of a transition from donor to acceptor is given by \cite{sinayskiy2012decoherence}
\begin{equation}
\label{eq-18}
\text{Max}[P_{D \to A}(t)] = \frac{\Delta^2}{\Delta^2 + \epsilon^2}
\end{equation}
where $\epsilon=\epsilon_D - \epsilon_A$ When the energy of the donor and the acceptor are equal, i.e., $\epsilon_D = \epsilon_A$, the maximum transfer probability from donor to acceptor $[P_{D\to A}(t)]$ at time $t_0= \pi/2 \Delta$ is equal to 1.
The Hamiltonian of the ligand, in this case the spike protein, is represented as a harmonic oscillator with frequencies $\omega$ associated with the protein:
\begin{equation}
\label{eq-19}
H_{P} =  \omega (b^{\dagger}b+1/2)
\end{equation}
\begin{equation}
\label{eq-20}
H_{R-P} = \sigma_z\sum_{i}  \omega_i \gamma_i(b + b^{\dagger}) \end{equation}

The interaction hamiltonain  extends to include both the donor and acceptor alongside the ligand protein, yet for computational simplicity, the latter's interactions are minimized in the numerical simulation. The interaction strength between the ligand protein and the receptor is symbolized by the coupling constant $\gamma$. The vibrational states of the spike protein are represented by creation and annihilation operators $b^\dagger$ and $b$, respectively, which correspond to a specific vibrational mode  $\omega$. The Hamiltonians $H_R$, $H_P$, and $H_{R-P}$ comprehensively describe our system, focusing on the dynamics of electron tunnelling within the context of SARS-CoV-2 receptor interactions.
\begin{equation}
\label{eq-21}
    H_{sys}=\frac{1}{2} \epsilon \sigma_z + \frac{1}{2} \Delta \sigma_x + \omega (b^{\dagger}b+1/2) + \sigma_z\sum_{i} \omega_i \gamma_i(b_i + b^{\dagger}_i)
\end{equation}
Similarly, the membrane environment is characterized by $H_E$, and its interplay with the receptor is approximated as $H_{R-E}$.
\begin{equation}
\label{eq-22}
H_E = \sum_{E}\omega_E(a_E^\dagger a_E + \frac{1}{2})
\end{equation}
\begin{equation}
\label{eq-23}
H_{R-E} = \sigma_z\sum_{i,E} {\omega_E}  \gamma_{iE} (a_E + a_E^\dagger)
\end{equation}

In the above equation, the term $\gamma_{iE}$ signifies the interaction strength between the receptor and its surrounding membrane environment, which is notably weaker compared to the interaction with the spike protein. The Hamiltonians $H_{R-P}$ and $H_{R-E}$ delineate distinct interactions; $H_{R-P}$ is essential for enabling the receptor's recognition of the spike protein, whereas $H_{R-E}$ encompasses a general coupling with various environmental vibrational mode  s adjacent to the receptor. This model discriminates between spike protein and environmental vibrations based on their unique frequencies and the relative strengths of their coupling constants. The summation in the equation accounts for interactions involving the donor, the acceptor, and all relevant environmental vibrations, reflecting the complex interplay within the system.

In our problem, the system is composed of the spike protein of the virus and the ACE2 receptor. The Hamiltonian of the system consists of $H_R$, $H_P$, $H_R-P$, and the Lindblad operator is given as $L=\gamma_{iE} \sigma_z$. Therefore, our (Eq.~\ref{eq-11}) in the non-Markovian case will have the following form:

\begin{multline}
\label{eq-24}
\frac{d}{dt} \tilde{\psi}_t = 
-i H_{\text{sys}} \tilde{\psi}_t 
+ \gamma_{iE} (\sigma_z - \langle \sigma_z \rangle) \tilde{\psi}_t \tilde{Z}_t 
- g_0(t) \gamma^2_{iE} \big[ (\sigma_z - \langle \sigma_z \rangle) \sigma_z- \langle\sigma_z - \langle \sigma_z \rangle\rangle \big] \tilde{\psi}_t\\ +i g_1(t) \gamma^2_{iE} \sigma_y (\Delta + 2 \gamma_i \omega) \big[ (\sigma_z - \langle \sigma_z \rangle) - 2 \langle \sigma_z - \langle \sigma_z \rangle \rangle \big] \tilde{\psi}_t
\end{multline}

where
\begin{equation}
\label{eq-25}
{\tilde{Z}t}=Z_t+\int_0^t ds \alpha(t-s) \langle \sigma_z^\dagger\rangle{s}\end{equation}
The coefficients $g_0(t), g_1(t)$ are given by (Eq.~\ref{eq-14}) and (Eq.~\ref{eq-15}), respectively.

By simulating (Eq.~\ref{eq-24}) and obtaining the density matrix by ensemble average from (Eq.~\ref{eq-5}), we find a maximum probability associated with the vibrational mode   in the dimer system. The difference in probabilities with and without the vibrational mode   takes the following form\cite{sinayskiy2012decoherence}:

\begin{equation} 
\label{eq-26}
\Delta P = Max[P_{D\to A}(t)]_{with} - {Max}[P_{D \to A}(t)]_{without}\end{equation}
Furthermore, the time-dependent transfer rate from the donor to the acceptor site can be calculated from the populations using the density matrix as follows:
\begin{equation} 
\label{eq-27}
\kappa = -\frac{d}{dt} \ln \left(P(t)  \right)
\end{equation}
\begin{table}[ht]
\centering
\begin{tabular}{|l|l|l|l|l|l|l|l|}
\hline
     & $\epsilon_{A}$ - $\epsilon_{D}$ & $\Delta$  & $\gamma_i$ & $\gamma_{iE}$& $\omega_{1}$ & $\omega_{2}$ & $\omega_{3}$ \\
\hline
Parameter ranges & 500-1700 $cm^{-1}$ & 0.0001-0.1 $eV$  & 0-0.49 $eV$ & 0.005-0.025 $eV$& 836 $cm^{-1}$ & 1000 $cm^{-1}$ & 1240 $cm^{-1}$ \\
\hline
\end{tabular}
\caption{\label{tab:example} The parameters for the numerical simulations\cite{solov2012vibrationally, huang2021site}}
\end{table}
\begin{figure}[h!]
\centering
\includegraphics[width=18cm]{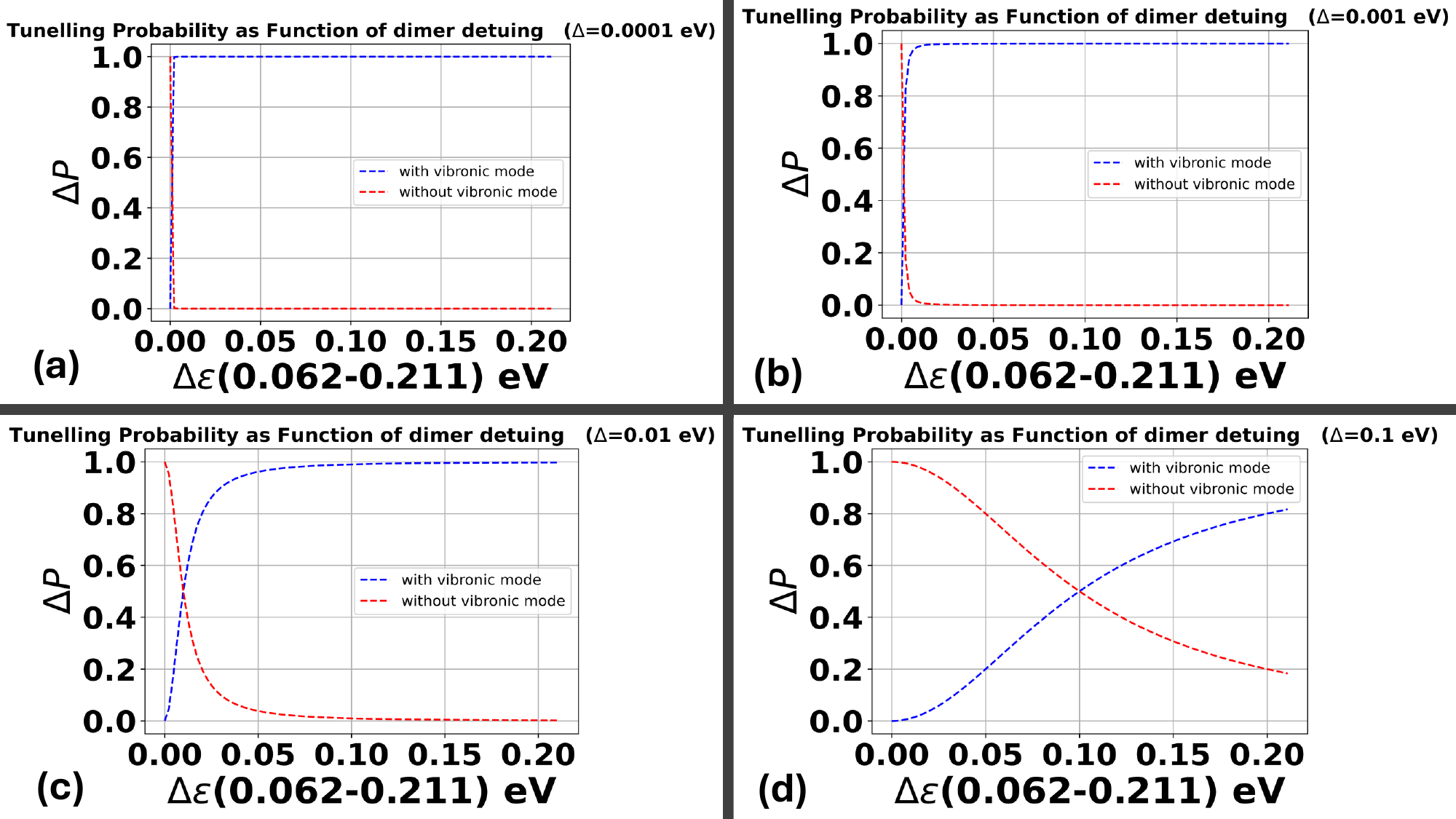}% Here is how to import EPS art
\caption{ (a-d) The figure presents a comparison of electron transfer probabilities with and without the spike protein as a function of \textit{dimer detuning}. The probabilities obtained without the presence of the spike protein are depicted by a \textbf{red dotted line}, exhibiting relatively uniform behavior across varying dimer detuning values.  In panel (a), corresponding to the \textit{weak coupling limit} (\(\Delta = 0.0001 \ \text{eV}\)), the influence of the vibrational mode on the transfer probability is minimal, resulting in negligible variation. Panel (b) shows the scenario at a slightly stronger coupling (\(\Delta = 0.001 \ \text{eV}\)), where a modest increase in transfer probability is observed. Panel (c) illustrates the case at \(\Delta = 0.01 \ \text{eV}\), where the effect of the vibrational mode on the transfer probability becomes more pronounced, indicating enhanced interaction between the system components. Finally, panel (d) represents the \textit{strong coupling limit} (\(\Delta = 0.1 \ \text{eV}\)), where a significant enhancement in transfer probability is observed due to the vibrational mode.The most striking feature highlighted in the (d) showing the positive probability at strong coupling, indicated by the blue dotted line.}
\label{fig-3}
\end{figure}

\section*{Results and Discussion}
 We have calculated  the max probability of electron transfer from the donor to the acceptor level, dependent on the dimer's detuning and coupling parameters. Using a non-Makovian Stochastic Schrödinger equation (Eq.~\ref{eq-24}) with Ornstein-Uhlenbeck noise and calculating the ensemble-averaged density matrix, we have formulated an expression that captures the maximum probability of this electron transfer. To effectively present these results, we constructed a 2D histogram that compares the maximum probabilities with and without the influence of the vibrational mode  , expressed as $\Delta P = \text{Max}[P_{D\to A}(t)]{\text{with  }} - \text{Max}[P{D \to A}(t)]_{\text{without  }}$.\\
 
We have gathered essential parameters from a range of biological processes to inform our analysis, notably adopting models from Solov’yov et al.\cite{solov2012vibrationally, huang2021site} concerning vibration-assisted tunnelling in olfactory receptors. These parameters are detailed in Table 1.

\begin{figure}[hbt!]
\begin{subfigure}{.475\linewidth}
  \includegraphics[width=\linewidth]{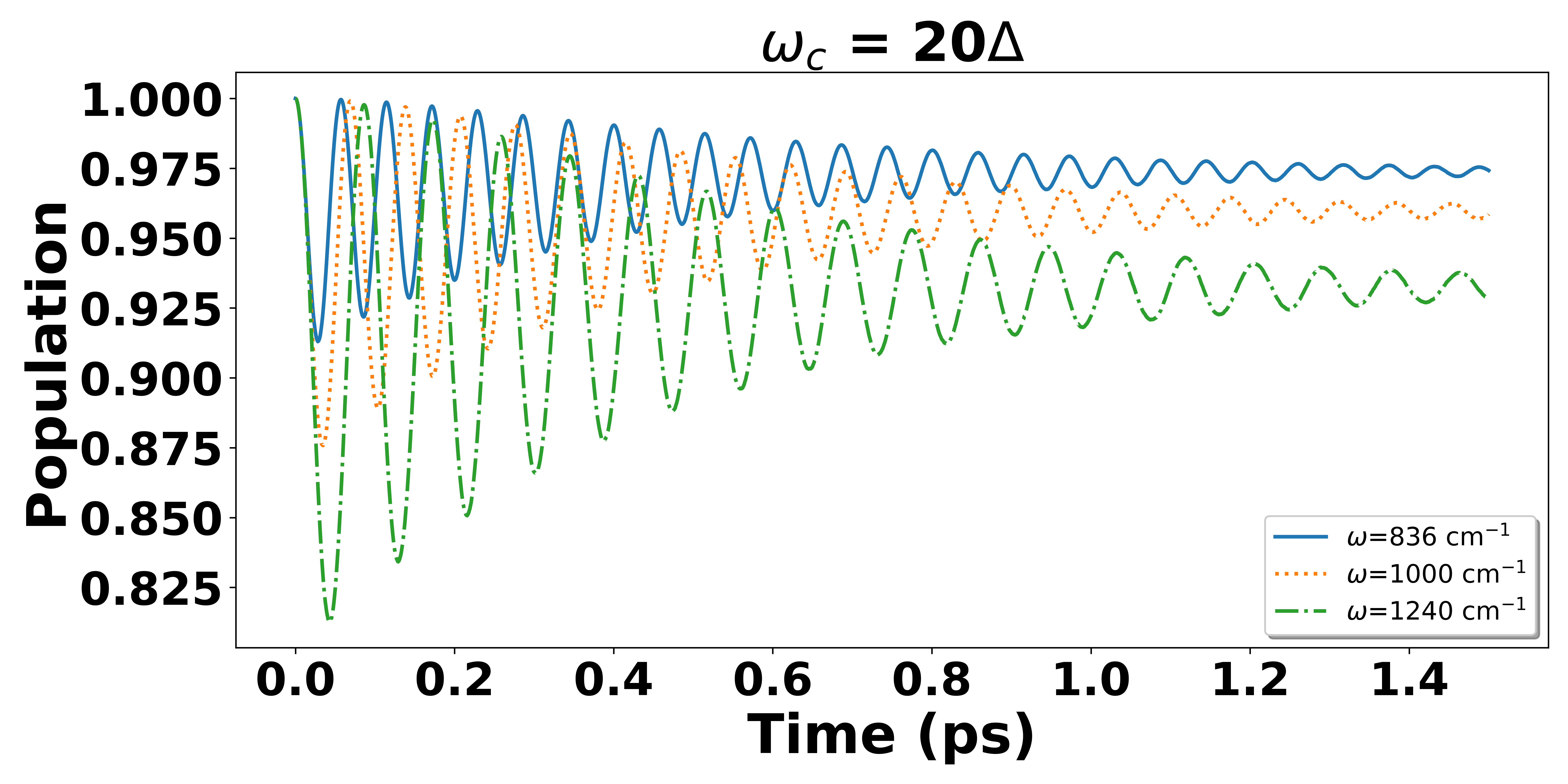}
  \caption{}
  \label{fig_population_sp_1}
\end{subfigure}\hfill % <-- "\hfill"
~ % optional tilde b/t figures for readability.
  % this solution will not work w/ empty lines b/t subfigures
\begin{subfigure}{.475\linewidth}
  \includegraphics[width=\linewidth]{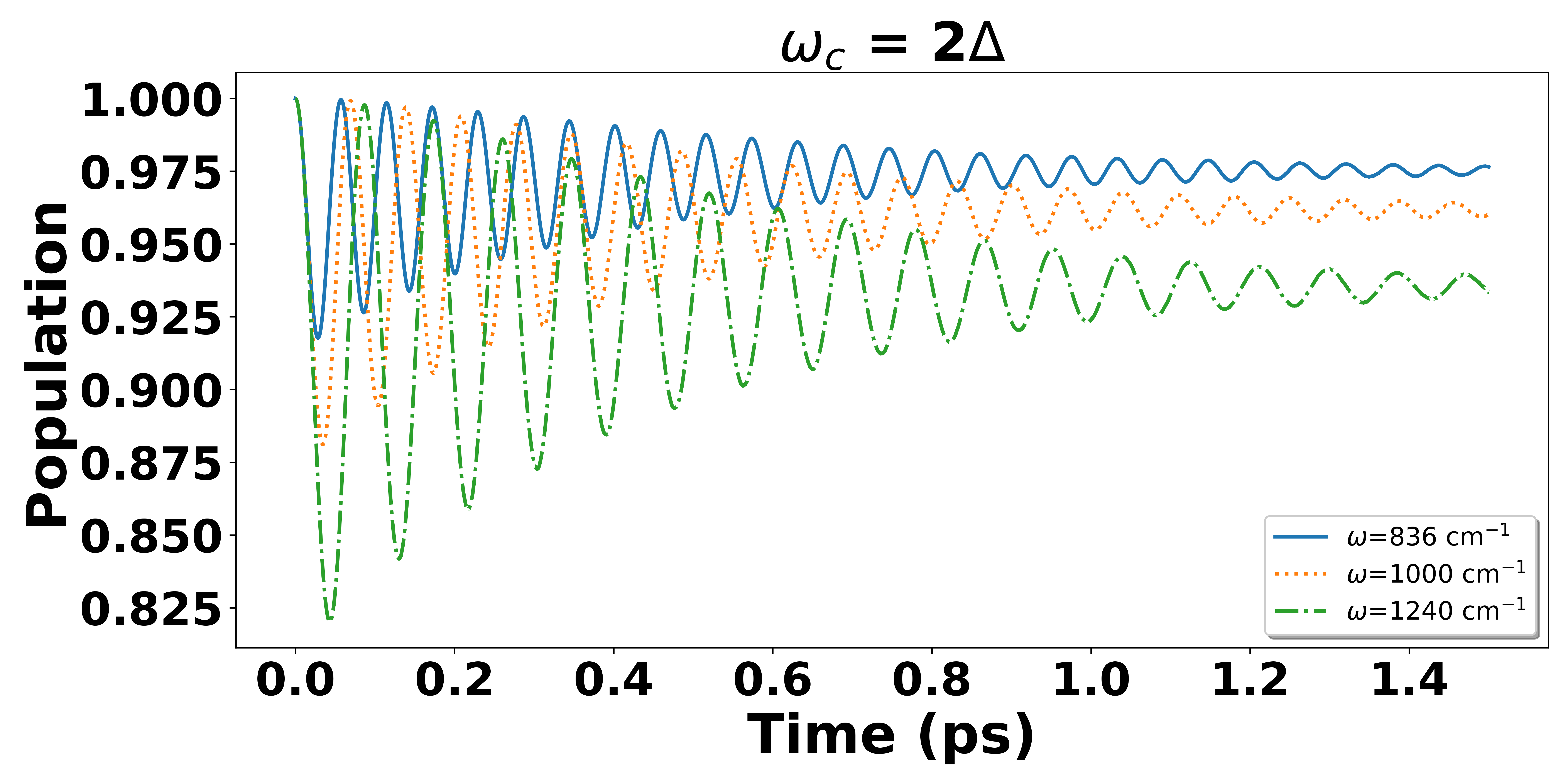}
  \caption{}
  \label{fig_population_sp_2}
\end{subfigure}
\medskip % create some *vertical* separation between the graphs
\begin{subfigure}{.475\linewidth}
  \includegraphics[width=\linewidth]{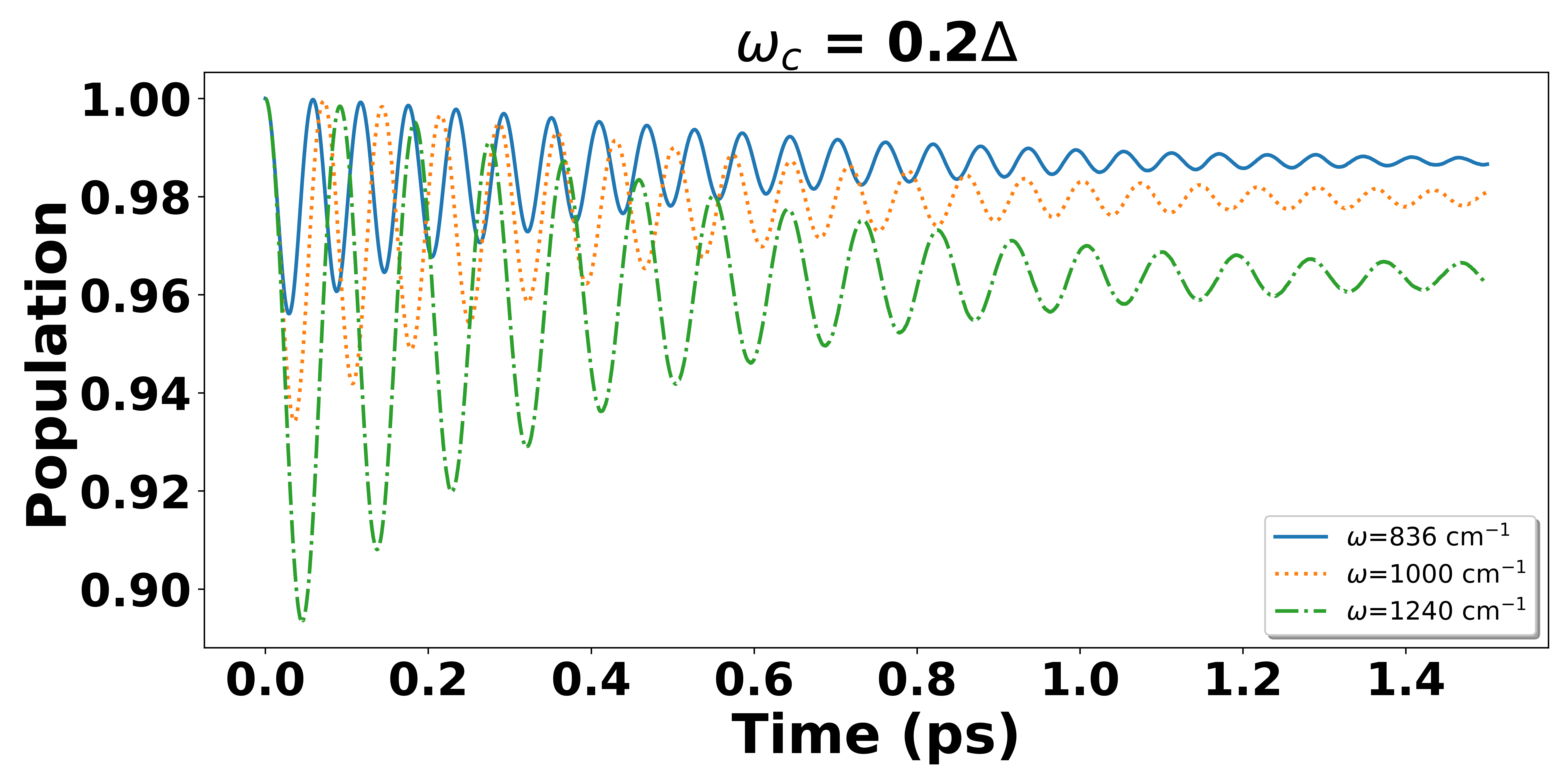}
  \caption{}
  \label{fig_population_sp_3}
\end{subfigure}\hfill % <-- "\hfill"
% or just a comment and no tilde
\begin{subfigure}{.475\linewidth}
  \includegraphics[width=\linewidth]{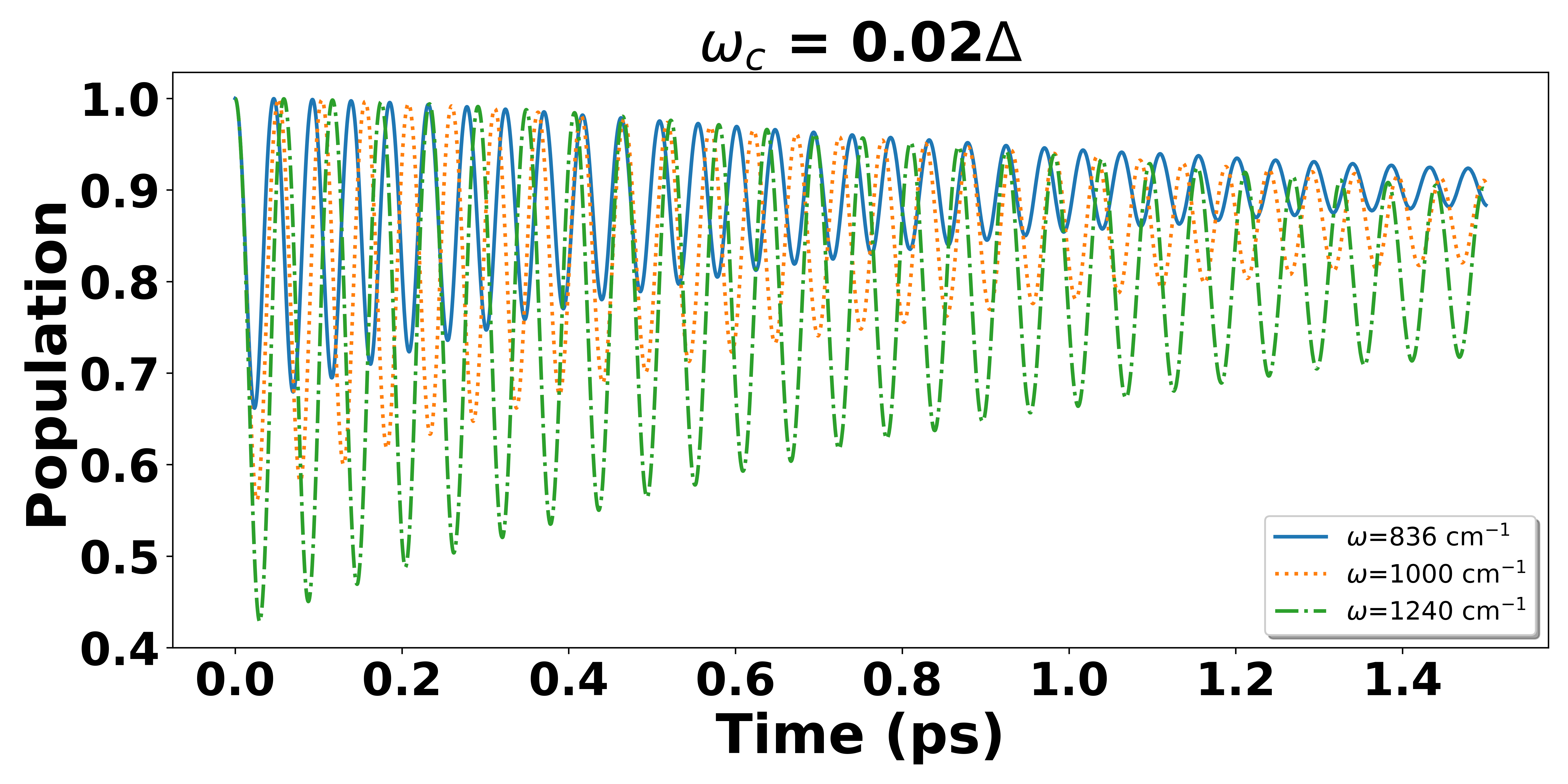}
  \caption{}
  \label{fig_population_sp_4}
\end{subfigure}
\caption{ Population dynamics simulated using the non-Markovian Stochastic Schrödinger Equation across different coupling strengths, illustrated in panels (a-d). (a) At $\omega_c = 20\Delta$, the weak coupling limit demonstrates a negligible impact of the vibrational mode on electron transfer probabilities. (b) Increasing to $\omega_c = 2\Delta$, there is a slight enhancement in transfer probabilities, indicating a minor influence of the vibrational mode. (c) At $\omega_c = 0.2\Delta$, intermediate coupling strengths show a more significant influence of the vibrational mode on transfer probabilities. (d) In the strong coupling limit at $\omega_c = 0.02\Delta$, a substantial effect of the vibrational mode is observed, markedly altering the population dynamics}. 
\label{fig:population}
\end{figure}

Our study draws significant parallels with the mechanism of olfaction, particularly in how electron transfer is facilitated. Normally, the energy difference $\epsilon_{A} - \epsilon_{D}$ between the redox potentials at the donor (D) and acceptor (A) sites is too substantial to allow for thermally assisted electron transfer within a biologically relevant time frame. However, binding of an appropriate odorant, which possesses a vibrational mode  matching $\epsilon_{A} - \epsilon_{D}$, permits this transfer by enabling the release of this energy difference into the odorant’s vibrational mode. These high-frequency modes are crucial for enabling electron transfer. Our model, inspired by such biological phenomena, applies similar principles to our case of infection, ensuring that the chosen parameters, including the tunnelling term $\Delta$, are biologically pertinent and reflective of physical processes observed in olfaction, thereby lending our theoretical framework greater biological accuracy and relevance\cite{chkecinska2015dissipation}.\\
In our study, we analyze the influence of vibrational frequencies on electron transfer between the spike protein and its receptor across varying coupling strengths, with dimer couplings set at $\Delta$ = 0.0001, 0.001, 0.01, and 0.1 eV \cite{chkecinska2015dissipation}, and the vibrational modes  informed by Raman spectroscopy research on SARS-CoV-2~\cite{huang2021site}. Initially, at a weak coupling limit of $\Delta$ = $0.0001 eV$, vibrational mode  s exhibit minimal impact on transfer probability, as shown in Figure~\ref{fig-2}(a), suggesting that the weak donor-acceptor coupling is insufficient to significantly facilitate electron tunnelling despite strong coupling between the spike protein and the receptor. As coupling strength increases to $\Delta$ = $0.001 eV$ Figure~\ref{fig-2}(b), there is a slight enhancement in transfer probability, indicating a marginal influence of the vibrational mode  .

\begin{figure}[hbt!]

\begin{subfigure}{.475\linewidth}
  \includegraphics[width=\linewidth]{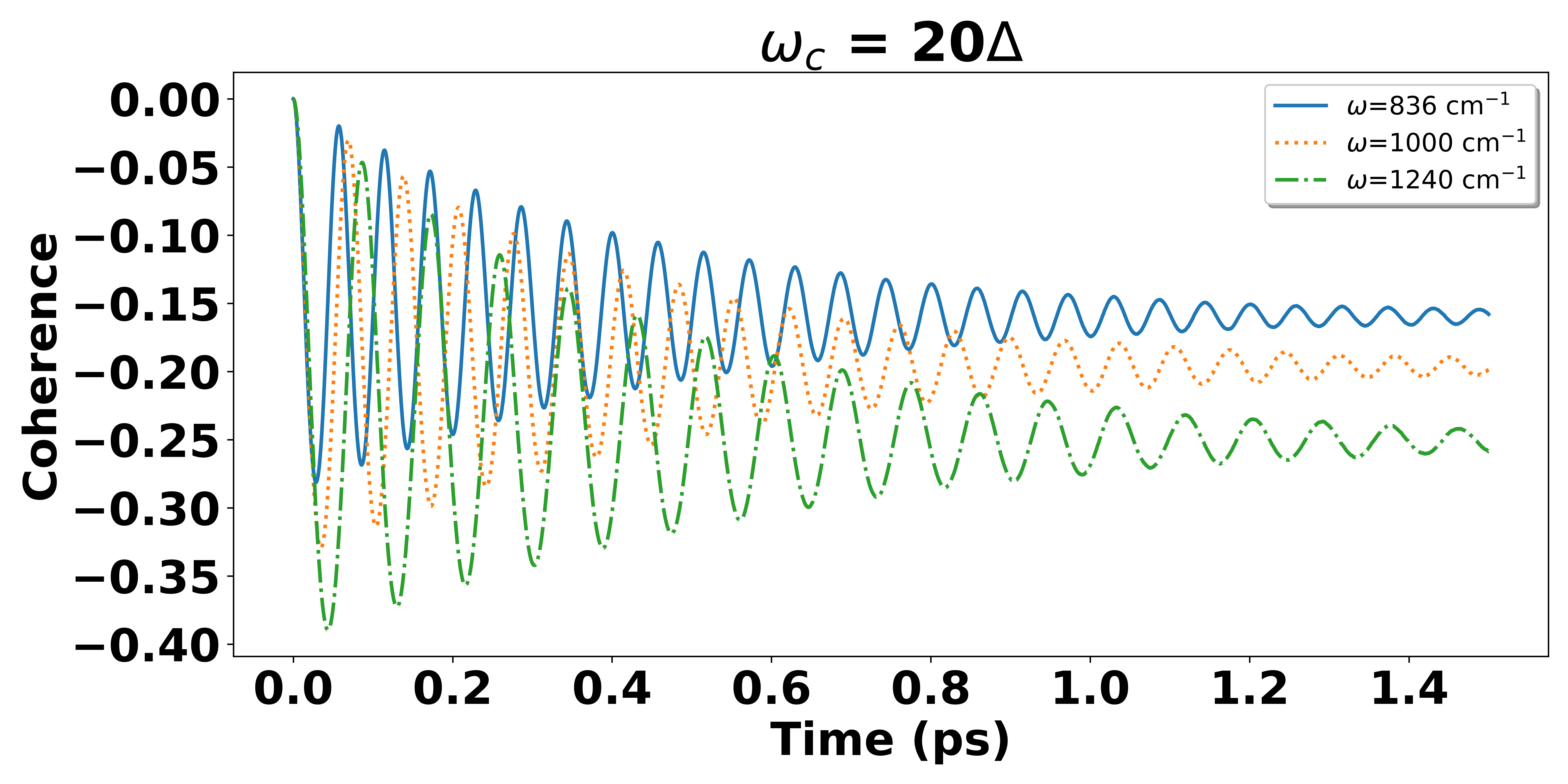}
  \caption{}
  \label{fig_Coherence_sp_1}
\end{subfigure}\hfill % <-- "\hfill"
~ % optional tilde b/t figures for readability.
  % this solution will not work w/ empty lines b/t subfigures
\begin{subfigure}{.475\linewidth}
  \includegraphics[width=\linewidth]{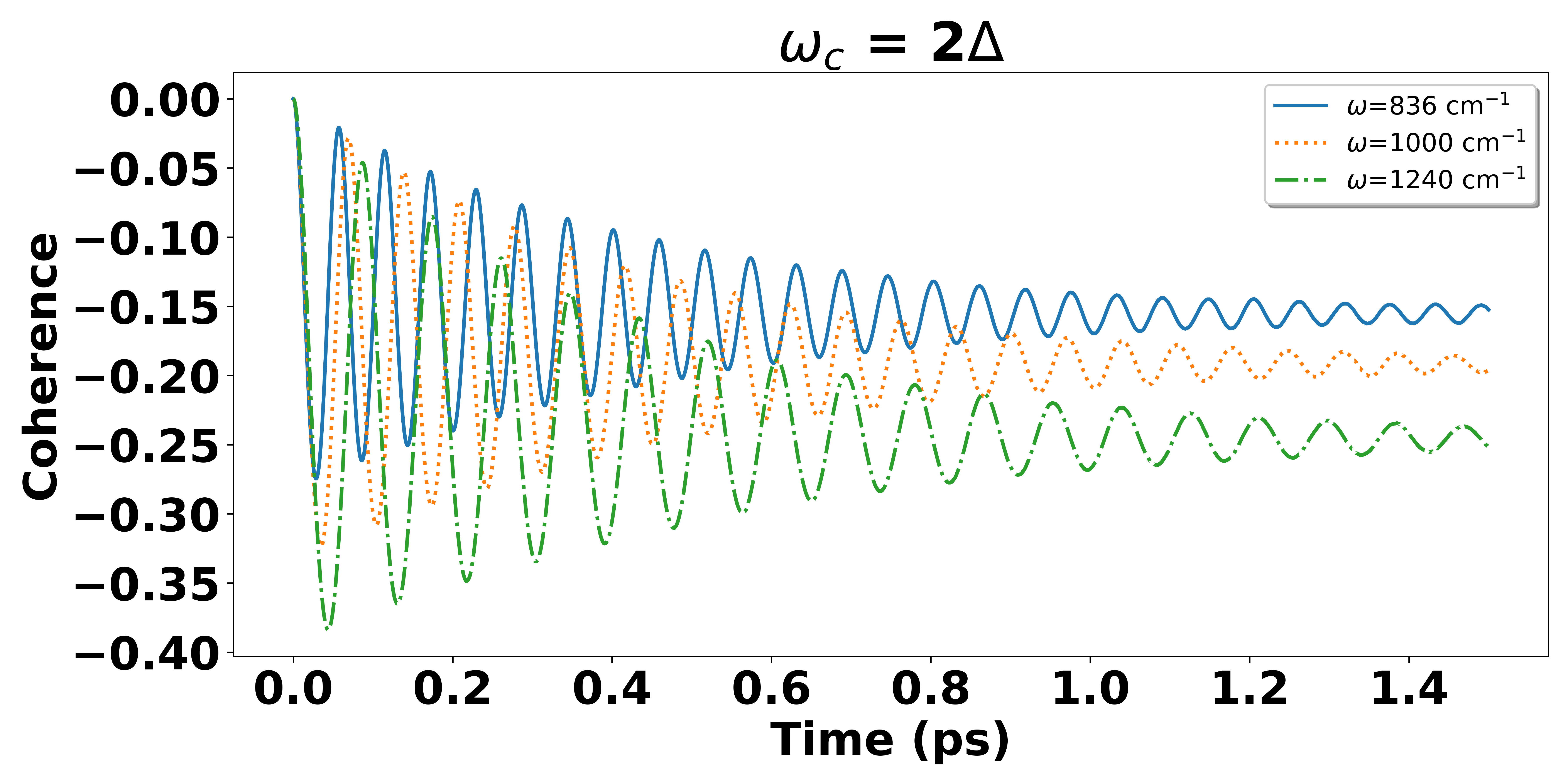}
  \caption{}
  \label{fig_coherence_sp_2}
\end{subfigure}

\medskip % create some *vertical* separation between the graphs
\begin{subfigure}{.475\linewidth}
  \includegraphics[width=\linewidth]{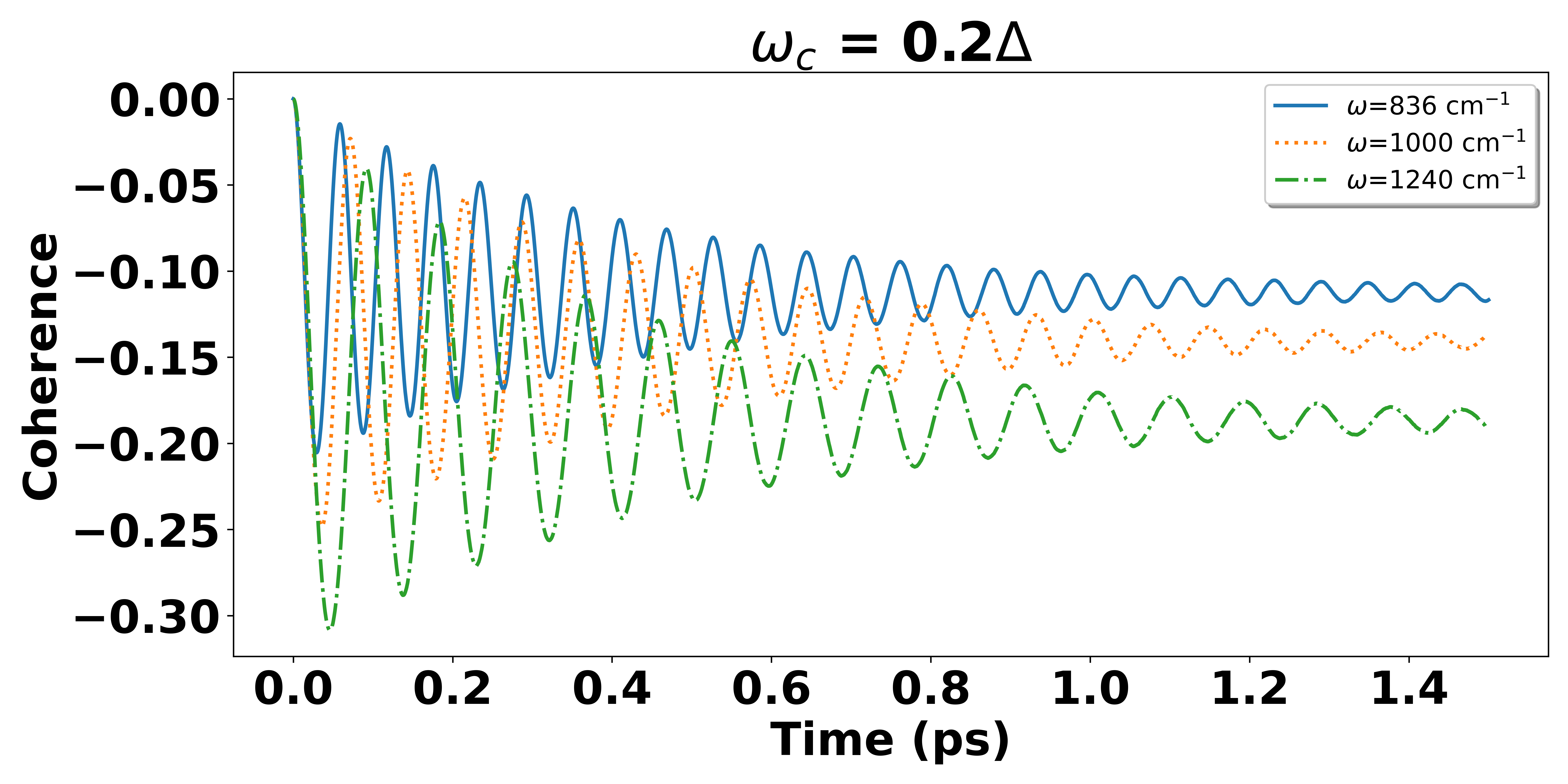}
  \caption{}
  \label{fig_Coherence_sp_3}
\end{subfigure}\hfill % <-- "\hfill"
% or just a comment and no tilde
\begin{subfigure}{.475\linewidth}
  \includegraphics[width=\linewidth]{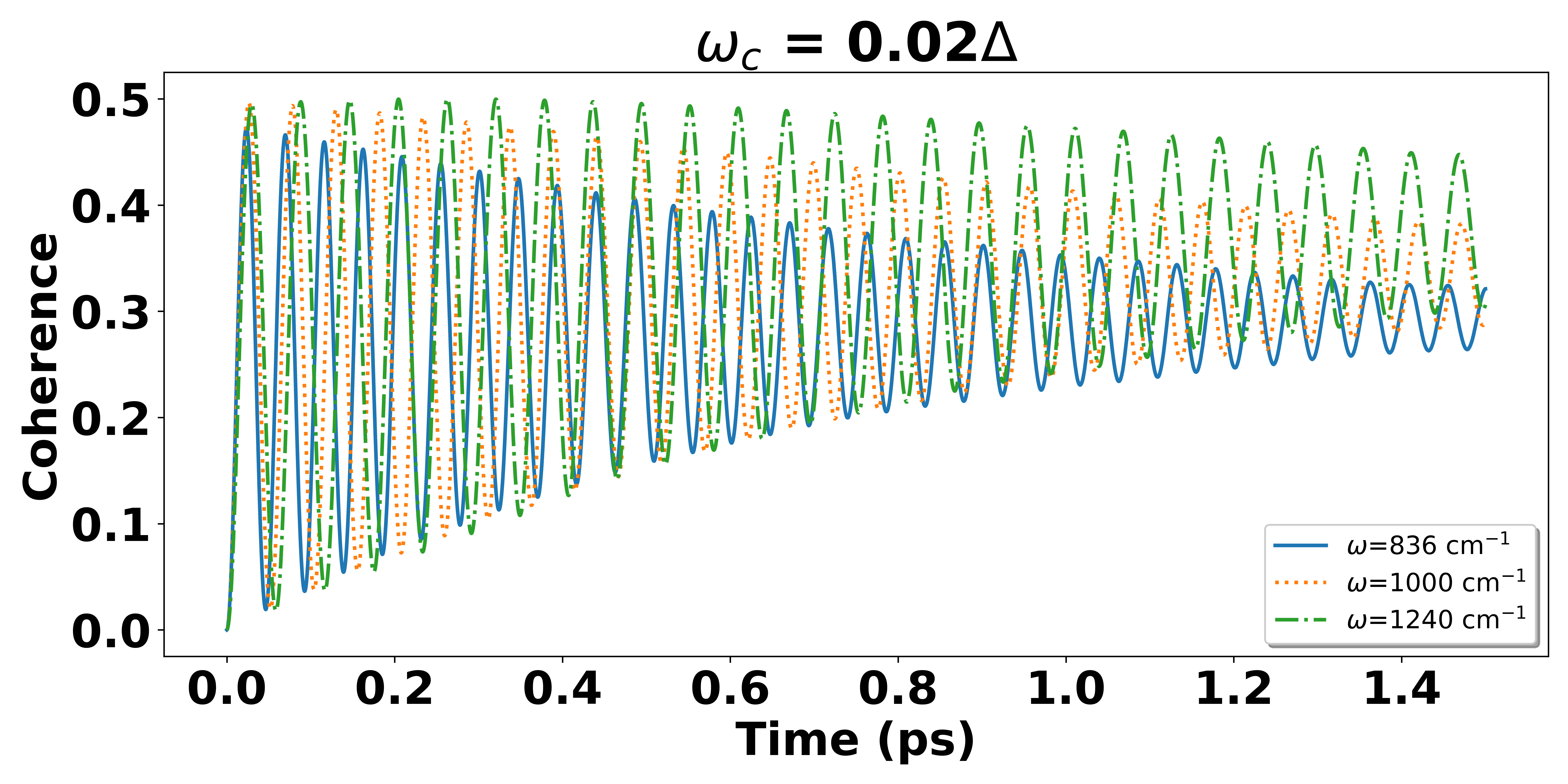}
  \caption{}
  \label{fig_coherence_sp_4}
\end{subfigure}
\caption{ Coherence dynamics simulated using the non-Markovian Stochastic Schrödinger Equation across various coupling strengths, depicted in panels (a-d). (a) At a weak coupling limit of $\omega_c = 20\Delta$, minimal impact on coherence dynamics is observed, with predominantly negative coherences. (b) At $\omega_c = 2\Delta$, there is no significant enhancement in coherence dynamics, indicating stable but unaltered states. (c) Intermediate coupling strength at $\omega_c = 0.2\Delta$ demonstrates a positive trend in coherence dynamics, suggesting increased interaction effects. (d) The strong coupling limit at $\omega_c = 0.02\Delta$ shows a substantial influence of the vibrational mode, significantly altering both population and coherence dynamics, highlighting the profound impact of stronger coupling and vibrational interactions on the system’s quantum coherence}.
\label{fig:Coherence}
\end{figure}

\begin{figure}[ht!]
\centering
\includegraphics[width=15cm]{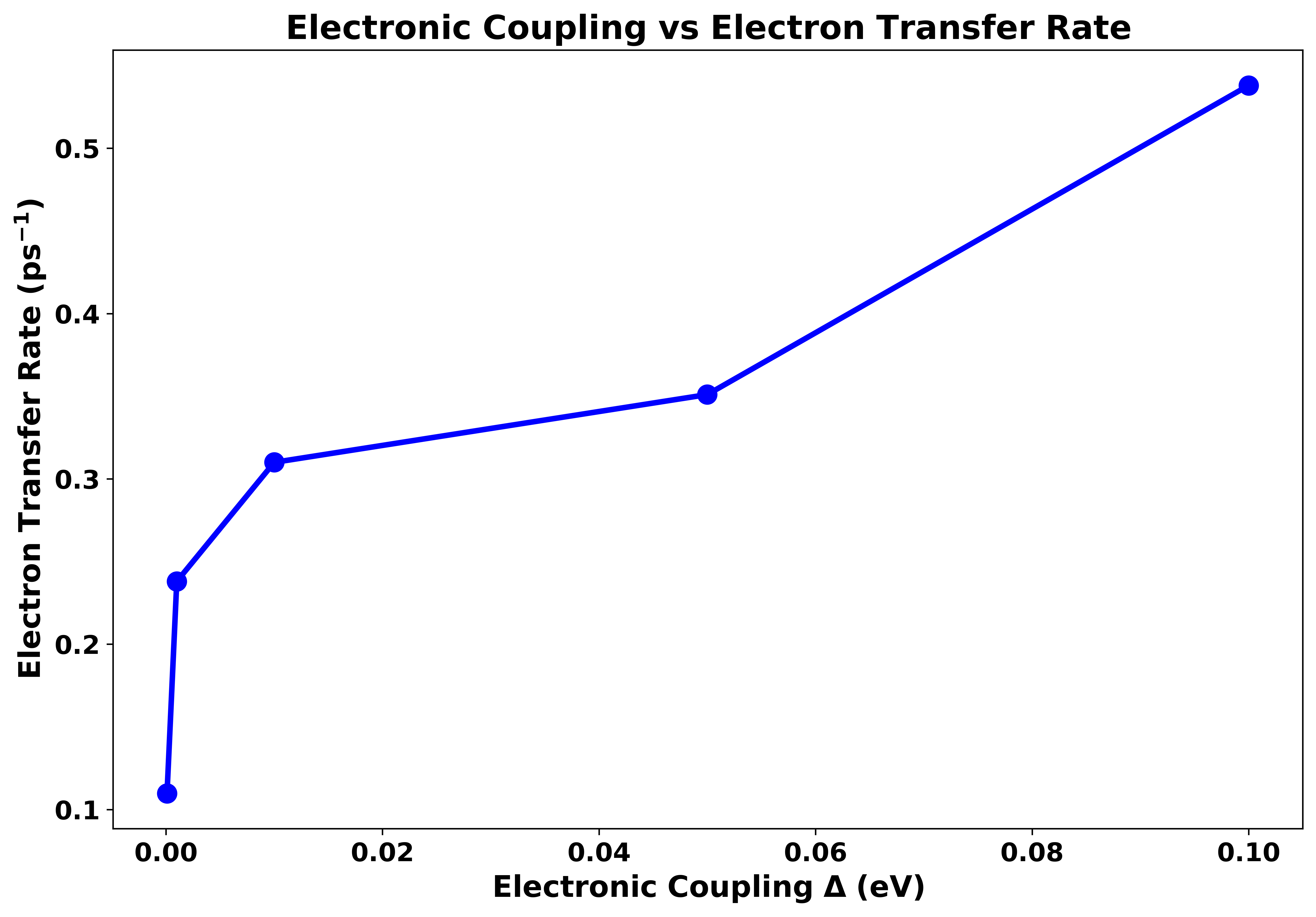}% Here is how to import EPS art
\caption{The figure shows the time-dependent decay rate (in ps$^{-1}$) as a function of the donor-acceptor coupling strength ($\Delta$) in the non-Markovian regime. The decay rate, calculated from the donor population dynamics, exhibits non-exponential behavior, characterized by transient negative values. 
These negative decay rates indicate population backflow and memory effects, which are hallmark features of non-Markovian dynamics. Such behavior demonstrates the reversible nature of electron transfer influenced by the environmental memory, highlighting the significance of non-Markovian effects in accurately modeling electron tunnelling processes, especially in the context of virus-host interactions.}
\label{fig-6}
\end{figure}

Further elevation of the coupling strength to $\Delta = 0.01 eV $Figure~\ref{fig-2}(c) marks a more substantial increase in electron tunnelling and transfer probability, suggesting that vibration-assisted electron tunnelling becomes progressively significant. At the highest examined strength $\Delta$ = $0.1 eV$, as shown in Figure~\ref{fig-2}(d)), the vibrational mode   significantly enhances the transfer probability, indicating a regime where strong coupling synergizes with vibronic assistance to markedly augment electron transfer. This observation underscores the critical role of coupling strength in modulating the interaction between the spike protein and the ACE2 receptor, which, in turn, influences the microscopic dynamics of virus infection.  This detailed understanding is pivotal for developing strategies to mitigate viral entry and infection.The process of electron transfer within molecular systems serves as a pivotal mechanism for molecular recognition, particularly through the detection of vibrational spectra associated with the virus's spike protein. This phenomenon, elucidated in reference \cite{chkecinska2015dissipation}, underscores the sensitivity of electron transfer rates to the presence or absence of odorant molecules, highlighting the intricate interplay between basic electron transfer processes and molecular interactions.This stands in stark contrast to the findings reported in previous studies, specifically within the Markovian regime, where extremely strong coupling between the donor and acceptor levels resulted in negative probabilities, which is physically implausible. \cite{adams2022quantum,haseeb2024vibration}\\ To further corroborate our findings on the non-Markovian nature of electron transfer in virus infection dynamics, Figure~\ref{fig-3}(a-d) displays the transfer probabilities with and without the influence of the vibrational mode  . Notably, the occurrence of positive probabilities at the strongest coupling strength, $\Delta = 0.1$ eV, underscores the distinctly non-Markovian behavior of the electron tunnelling process.\\
We have broadened our analysis to encompass the populations and coherences across three distinct vibrionic modes, spanning a range from weak to strong coupling between the donor and acceptor levels in relation to the cutoff frequency, $\omega_c$. Figure~\ref{fig:population},~\ref{fig:Coherence} delineates the populations and coherences for the three specific vibrionic modes, $\omega= 836, 1040, 1240  cm^{-1}$.
Figure~\ref{fig_population_sp_1},~\ref{fig_Coherence_sp_1} illustrates the populations and coherences at $\Delta=20 \omega_c$. Consistent with previous observations in Figures~\ref{fig-1}, where the electron transfer probability was minimal, a similar trend is noticeable here, with the graphs exhibiting low population levels and negative coherence. As the detuning increases to $\Delta=2 \omega_c$ Figure~\ref{fig_population_sp_2},~\ref{fig_coherence_sp_2}, a marginal rise in the transfer probability is observed, suggesting a slight enhancement in electron mobility.
Moving to Figure~\ref{fig_population_sp_3},~\ref{fig_Coherence_sp_3}, which represents the scenario at $\Delta=0.2 \omega_c$, it is apparent that the transfer probability begins to escalate from this point, with coherence also adopting a positive character. The most significant enhancement in transfer probability is evident in Figure~\ref{fig_population_sp_4},~\ref{fig_coherence_sp_4} at $\Delta=0.02 \omega_c$. Additionally, with the increased value of the vibrionic mode, more pronounced oscillatory behavior is observed in both the populations and coherences, indicating a more dynamic interplay between these quantum states.\\
Furthermore, we further extend our analysis by calculating the time-dependent decay rate from the donor to the acceptor site. Figure~\ref{fig-6} presents the time-dependent decay rates (in ps$^{-1}$) as a function of the donor-acceptor coupling strength ($\Delta$). The results clearly illustrate the \textit{non-exponential decay rate}, which is a hallmark of \textit{non-Markovian dynamics} in the viral infection process. In contrast, within the \textit{Markovian regime}, the decay rates increase linearly with the coupling strength, reflecting the loss of memory effects typical of Markovian behavior. These findings highlight the importance of incorporating \textit{non-Markovian effects} when modeling electron transfer processes in virus-host interactions, as they more accurately capture the dynamical characteristics of the system compared to conventional Markovian approaches.

\textcolor{black}{Although there is no  experimental study specifically focusing on testing the  mechanisms of virus infection at molecular level. our review of the literature highlights a potential technique that could experimentally validate electron tunneling mechanisms at the molecular level: Quantum Biological Electron tunnelling (QBET) spectroscopy. QBET spectroscopy offers the capability to optically detect quantum electron tunnelling in real time, allowing for the visualization of electron transfer (ET) dynamics in biological systems, such as enzymes. This method, as demonstrated in a study involving mitochondrial cytochrome c, has proven effective in capturing the spatiotemporal behavior of electron tunnelling in live biological environments\cite{xin2019quantum}.
In the case of SARS-CoV-2, QBET could be adapted to investigate the electron transfer processes between the spike protein and the ACE2 receptor, with particular emphasis on the involvement of disulfide bridges and cysteine residues that may facilitate redox interactions. By conjugating plasmonic nanoparticles to both the spike protein and the ACE2 receptor, QBET spectroscopy could detect quantized electron transfer events by monitoring changes in the optical scattering spectrum in real time. This approach holds the potential to provide a deeper understanding of the role electron tunnelling may play in viral entry mechanisms.}
% \begin{figure}[h!]
% \centering
% \includegraphics[width=18cm]{Figure 5.png}% Here is how to import EPS art
% \caption{ Coherence dynamics simulated using the non-Markovian Stochastic Schrödinger Equation across various coupling strengths, depicted in panels (a-d). (a) At a weak coupling limit of $\omega_c = 0.02\Delta$, minimal impact on coherence dynamics is observed, with predominantly negative coherences. (b) At $\omega_c = 0.2\Delta$, there is no significant enhancement in coherence dynamics, indicating stable but unaltered states. (c) Intermediate coupling strength at $\omega_c = 2\Delta$ demonstrates a positive trend in coherence dynamics, suggesting increased interaction effects. (d) The strong coupling limit at $\omega_c = 20\Delta$ shows a substantial influence of the vibrational mode, significantly altering both population and coherence dynamics, highlighting the profound impact of stronger coupling and vibrational interactions on the system’s quantum coherence.}
% \label{fig-5}
% \end{figure}

\section*{Conclusions}
In this study, we have applied the quantum state diffusion methodology to address the dynamics of COVID-19 infection through the framework of the Non-Markovian Stochastic Schrödinger equation. Our investigation primarily focused on the within biologically relevant parameters and its enhancement of viral infection at the microscopic level. By conducting simulations of the Non-Markovian Stochastic Schrödinger equation and analysis of simulation results, we have uncovered significant insights into how vibration-assisted electron tunnelling influences the infection process. Specifically, our findings elucidate the dependency of this process on the coupling strength between the SARS-CoV-2 spike protein and the ACE2 receptor, offering a deeper understanding of the mechanistic aspects of viral entry into host cells.\\
A critical aspect of our research is the examination of non-Markovian dynamics, which are crucial for understanding the nuances of the electron tunnelling process during viral infections. Our study reveals that tunnelling probability is minimal under weak coupling, increasing significantly at intermediate and strong coupling strengths as the coupling reaches intermediate levels, becoming most pronounced at strong coupling. Moreover, our findings emphasize the influence of the vibrational mode; significant enhancements in population transfer and coherence dynamics are observed between the donor and acceptor levels, significant enhancements in population transfer and coherence dynamics are observed. Notably, by employing the same parameters used in the Markovian approximation, our results demonstrate that the tunnelling probability remains positive even in the strongest coupling limit—a stark contrast to predictions made as predicted by Markovian models. This highlights the importance of considering non-Markovian effects for a more accurate and comprehensive understanding of the quantum mechanical processes underlying viral infection dynamics.

\section*{Acknowledgments}
This work is supported by United Arab Emirates University  
\section*{Author Contributions Statement}
Muhammad Waqas Haseeb and Mohammad Toutounji made significant contributions to the research described in this manuscript. As the lead author, Muhammad Waqas Haseeb was tasked with drafting the initial manuscript, performing essential calculations and data analysis under the guidance of Mohammad Toutounji. Mohammad Toutounji played a pivotal role in shaping the study's conceptual framework, providing critical guidance and continuous support throughout the research and writing processes. He also thoroughly reviewed and refined the manuscript to ensure clarity and scientific rigor. Ultimately, both authors have carefully reviewed and given their final approval for the published version of the manuscript.
\section*{Data availability}
The research data and parameters referenced in this study are documented in the referenced publications. For further inquiries or access to the datasets, interested parties may contact the corresponding author, who will provide the data upon reasonable request.
\appendix
\bibstyle{unsrt}
\bibliography{sample}

\end{document}